\documentclass[letterpaper]{jpconf}
\usepackage{graphicx}
\usepackage{floatflt}
\usepackage{wrapfig}

\newcommand{\Lambar}{\overline{\Lambda}}

\newcommand{\ra}{\rightarrow}

\newcommand{\ksks}{$K_S^0 K_S^0$~}
\newcommand{\Ks}{$K_S^0$~}
\newcommand{\ksf}{\ensuremath{K^0_S}}
\newcommand{\ffa}{$f_{2}(1270)/a_{2}^{0}(1320)$~}
\newcommand{\ffb}{$f_{2}^{'}(1525)$~}
\newcommand{\ffc}{$f_{0}(1710)$~}

\newcommand{\coll}{Collaboration.~}

\begin{document}
\title{Particle production at HERA}

\author{Changyi Zhou\footnote{On behalf of the H1 and ZEUS collaborations}}

\address{McGill University, \em{czhou@physics.mcgill.ca}}

\begin{abstract}
H1 has measured a number of different known particles and compared
their production to QCD models and to other reactions such as N-N
collisions. ZEUS has also measured the production of \ksks pairs
with a view to searching for glueballs. Several resonances are seen
which are glueball candidates. The results on the masses and widths
are compared to other experiments.
\end{abstract}

At HERA acceleration ring, electron (positron) and proton beams were
accelerated and made to collide with a center-of-mass energy of
above 300 GeV at two
interaction points where the H1 and ZEUS detectors were located.\\

\section{Energy dependence of the charged multiplicity in deep inelastic scattering}
The production of multi-hadronic final states has been an
interesting subject. Charged multiplicity in DIS are measured in
both the hadronic center-of-mass (HCM) frame and the Breit frame,
using an integrated luminosity of $38.6~pb^{-1}$. Only hadrons in
the current fragmentation regions of both frames were used due to
detector acceptance limitations.

\begin{figure}[hbtp]

\includegraphics[width=0.5\textwidth]{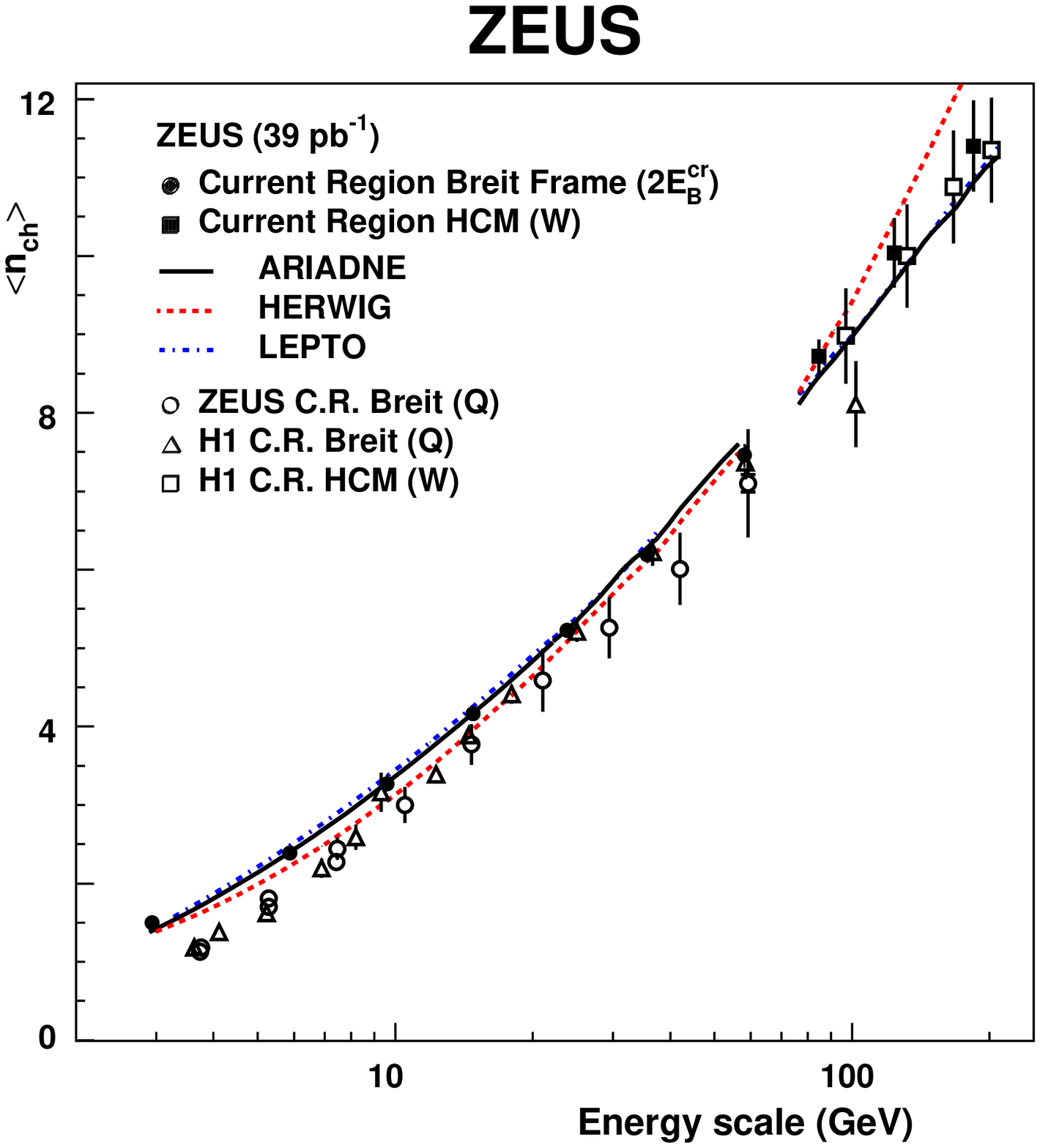}
\hfill
\includegraphics[width=0.5\textwidth]{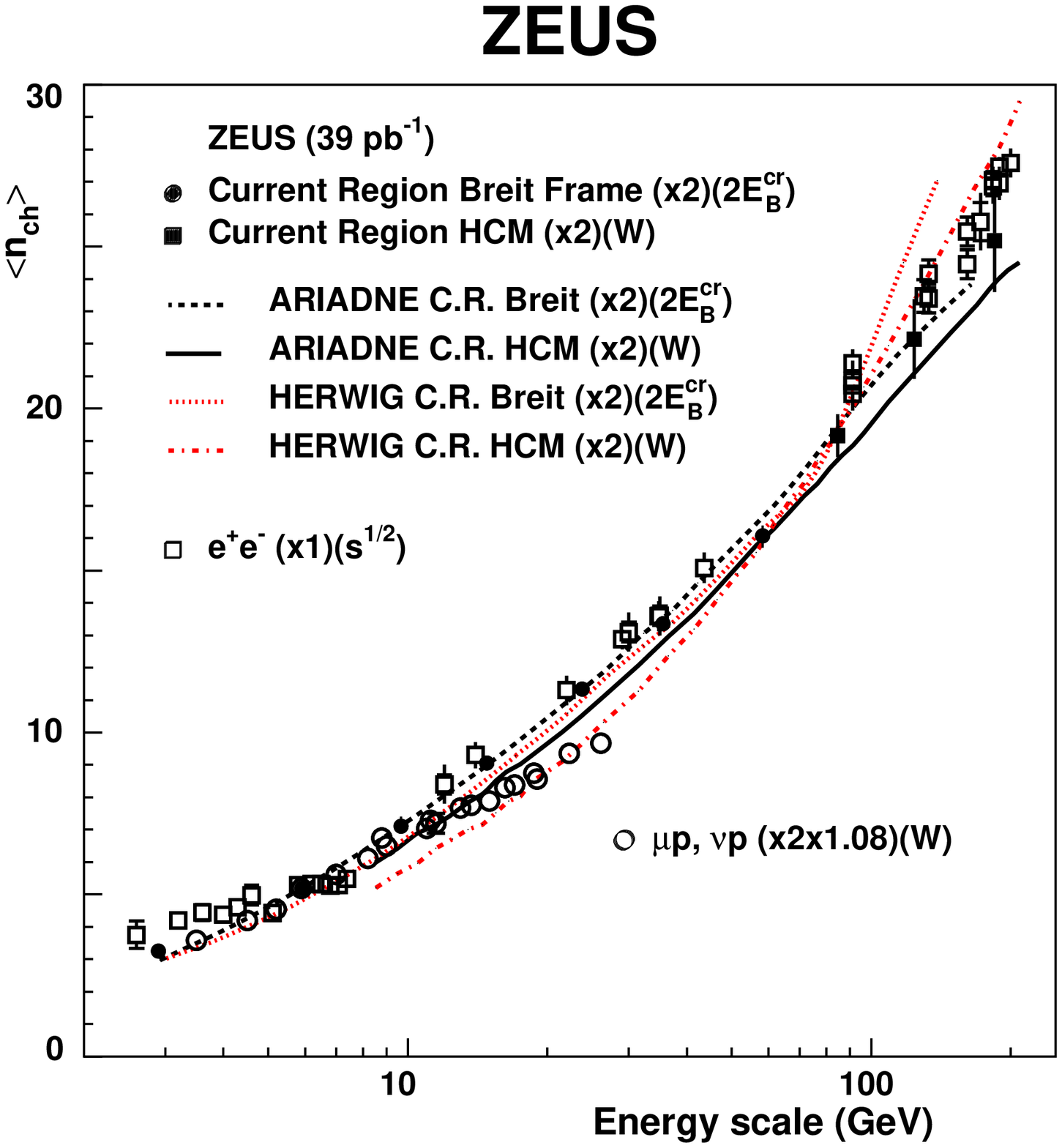}
     \caption{Mean charged multiplicity, $<n_{ch}>$, as a function of energy scale.}
    \label{multipli}

    \end{figure}

The mean charged multiplicity is shown in Figure
\ref{multipli}(left) in the current region of the HCM frame as a
function of energy, W, and in the current region of the Breit frame
as a function of $2 \cdot E^{cr}_B$. The data are compared and in
good agreement with the ARIADNE and LEPTO Monte Carlo predictions,
while results from HERWIG Monte Carlo prediction are lower. The data
are also compared with results of previously published
ZEUS~\cite{zfp:c67:93} and
H1~\cite{pl:b654:148,zfp:c72:573,np:b504:3} measurements. The
measurements agree within the experimental uncertainties at higher
values of $2 \cdot E^{cr}_B$, but differ at low values as a function
of $Q$. The energy scale $2 \cdot E^{cr}_B$ gives better agreement
in the current region of the Breit frame and results from $e^+e^-$
experiment than using $Q$. The mean charged multiplicities are
compared with results from
$e^+e^-$~\cite{pl:b70:120,zfp:c20:187,pr:d34:3304,zfp:c45:193,
zfp:c35:539,zfp:c50:185,pl:b273:181,zfp:c73:409,
pl:b372:172,pr:399:71,cern-ppe/96-47,pl:b416:233} and
fixed-target~\cite{zfp:c76:441,zfp:c35:335,zfp:c54:45} experiments
in Figure \ref{multipli} (right). Both the current regions of the
Breit and HCM frames are used as function of the energy scales, $2
\cdot E^{cr}_B$ and W, respectively. Good overall agreement is
observed among these experiments, presenting the same dependence of
the mean charged multiplicity on the respective energy scale. In
general conclusion, universality of mean charged multiplicity
dependence with energy scale is observed.

\section{Strangeness production at low $Q^2$ in deep inelastic $ep$ scattering}

The production measurement of strangeness production gives insight
into the suppression of strangeness involved in lighter flavours
fragmentation process studies. This non-perturbative colour string
fragmentation process dominates the mechanism of strange hadrons,
followed by the QPM(Quark Parton Model), BGF(Boson-Gluon Fusion) and
heavy quark decays.

Neutral strange particles (\Ks and $\Lambda$($\Lambar$)) production
in DIS with $2 < Q^2 < 100~GeV^2$ are measured by the H1
collaboration with the data sample collected in the years 1999 and
2000, superseding a previous H1 publication~\cite{H1-K0} with 40
times larger statistics.

The neutral strange meson \Ks and baryon $\Lambda$($\Lambar$) are
reconstructed via their dominant decay channels: $K^0_S \ra
\pi^+\pi^-$ and $\Lambda(\bar \Lambda) \ra p(\bar p)
  \pi^-(\pi^+)$. Various selection cuts are applied to purify the
  data sample.
  
The total inclusive cross section $\sigma_{\it vis}$ in the selected
kinematic region is given by:

\begin{equation}
\sigma_{vis}(ep \ra e [\ksf, \Lambda, h^{\pm}] X) = \frac{N}
  { {\cal L} \cdot  \epsilon \cdot BR \cdot (1 + \delta_{rad}) }\qquad
\label{cross-section}
\end{equation}

where $N$ represents the observed number of $K^0_s$, the sum of
$\Lambda$ and $\Lambar$ baryons or the charged hadrons $h^{\pm}$,
respectively. $\cal{L}$ is the integrated luminosity. $BR$ are the
branching ratios for the $K^0_s$ and $\Lambda$ decays and $BR=1$ for
charged hadrons.

\begin{figure}[hbtp]

\includegraphics[width=0.5\textwidth]{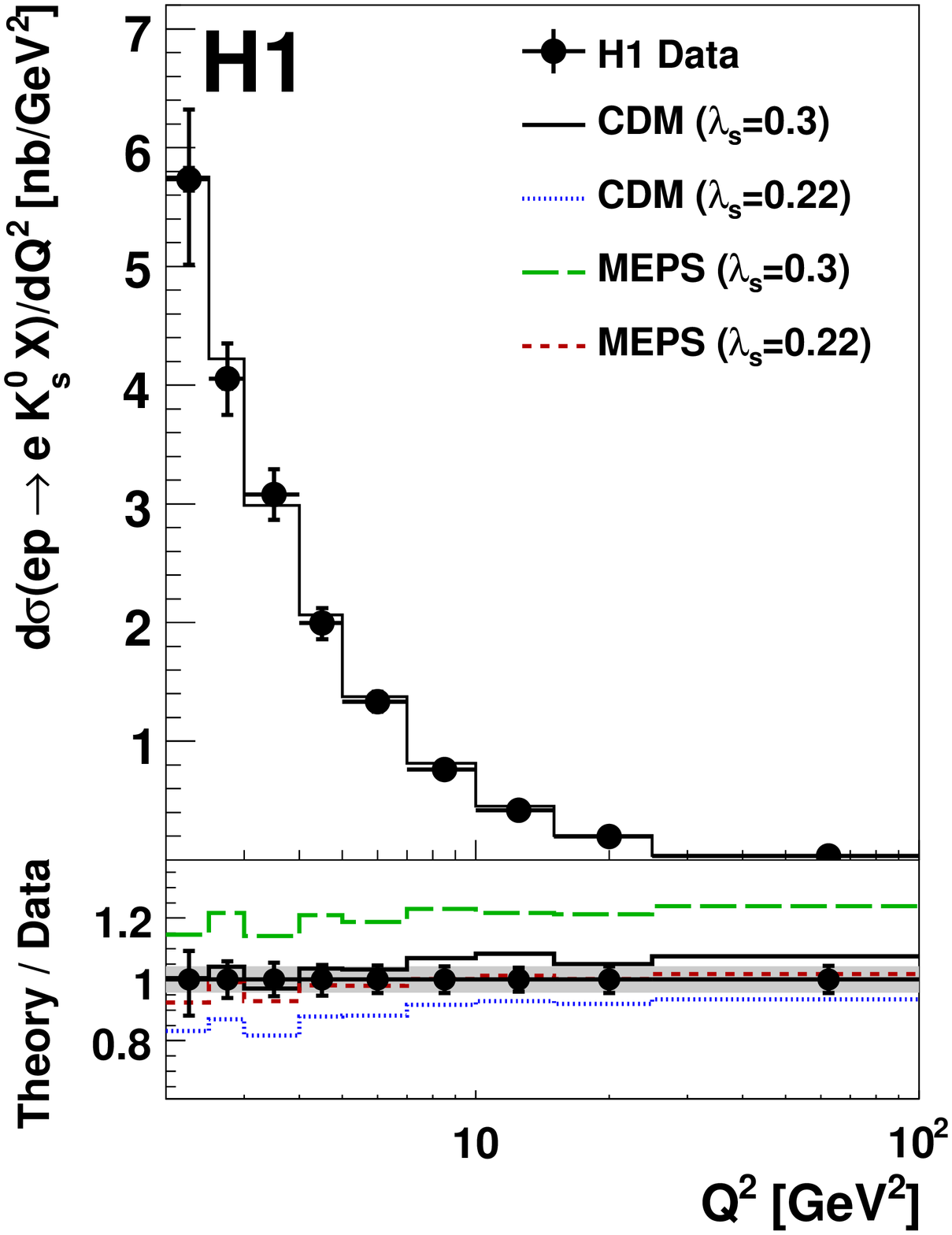}
\includegraphics[width=0.5\textwidth]{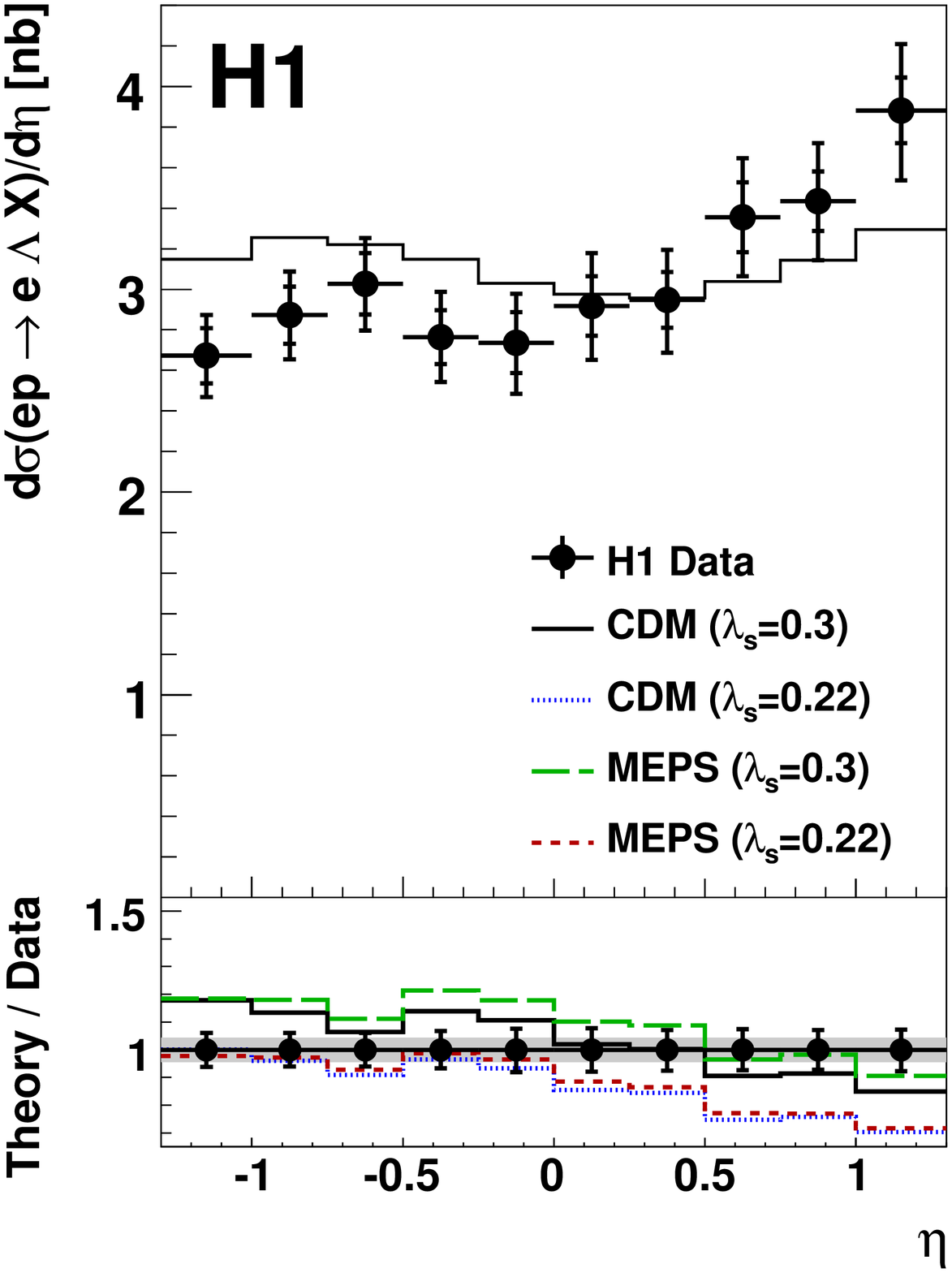}
 \caption{The differential cross section in the laboratory frame for \Ks as a function
 of the photon virtuality squared $Q^2$ (left) and for $\Lambda$ as
 a function of pseudorapidity $\eta$ (right).}
    \label{strange}

    \end{figure}

The production cross section and their ratios of \Ks, $\Lambda$ and
charged hadrons are measured inclusively. The results are in good
agreement with DJANGOH Monte Carlo predictions. The cross section
ratio is better described by the CDM(Colour Dipole Model) than the
MEPS(Matrix Element plus Parton Shower). The \Ks to charged hadrons
cross section ratio shows better agreement with the data with
$\lambda_s = 0.22$, where $\lambda_s$ is the ratio of the
possibilities of the strange and light quark productions. The result
is demonstrated in Figure~\ref{strange}.

\section{Inclusive photoproduction of $\rho^0$, $K^{\*0}$ and $\phi$ mesons}

Inclusive non-diffractive photoproduction of $\rho(770)^0$,
 $K^*(892)^0$ and $\phi(1020)$ mesons is investigated by the H1 collaboration with an integrated
 luminosity of ${\cal L}=36.5$~pb$^{-1}$ taken in year 2000.

\begin{wrapfigure}{r}{0.5\textwidth}
\begin{center}
\vspace{-0.8cm}
\includegraphics[width=0.5\textwidth]{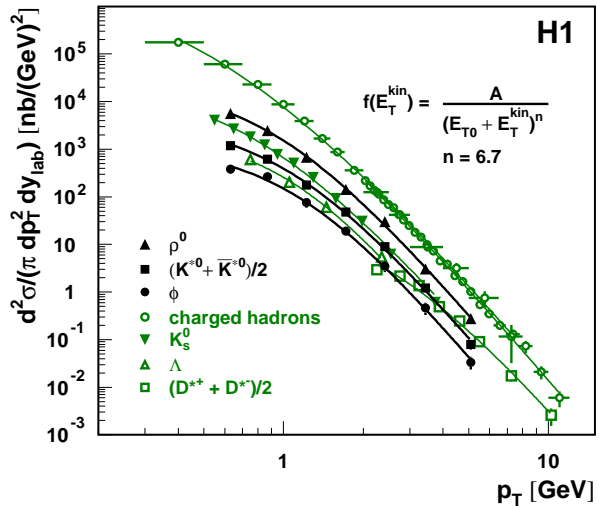}
\vspace{-0.8cm} \caption{The curves are the fits to the power law.}
 \label{rho}
 \vspace{-1.0cm}
 \end{center}
 \end{wrapfigure}

 The mesons are reconstructed using their dominant decay channels: $\rho(770)^0\to\pi^+\pi^-$,
$K^*(892)^0\to K^+\pi^-$ or $\overline{K}^*(892)^0\to K^-\pi^+$ and
$\phi(1020)\to K^+K^-$. Differential cross sections are presented as
a function of transverse momentum in Figure \ref{rho}, and are
compared to the predictions of hadroproduction models.

The results supports the thermodynamic picture of hadronic
interactions~\cite{therm}, in which the different primary hadrons
are thermalized and produced in different extends.

The cross section ratios are determined and compared to results
obtained in $pp$ and heavy-ion collisions. General agreements are
achieved.

\section{Inclusive \ksks resonance production in $ep$ collisions}
The lightest glueball is predicted by
theory~\cite{Klempt:2007cp,Oller} to have $J^{PC}=0^{++}$ and it has
a mass in the range 1450--1750 MeV. It can mix with $q\overline{q}$
states from the scalar meson nonet, like \ksks resonance states.

Inclusive \ksks production in $ep$ collisions at HERA has been
studied with the ZEUS detector using the full set of the integrated
luminosity of 0.5 $fb^{-1}$. \Ks mesons were reconstructed through
the charged-decay mode, $K_s^0 \to\pi^{+}\pi^{-}$. The \ksks
invariant mass distribution was reconstructed by combining two \Ks
candidates.
 Enhancements attributed to
 the production of \ffa, \ffb and \ffc are
 observed in the \ksks mass spectrum.

\begin{wrapfigure}{r}{0.5\textwidth}
\begin{center}
\vspace{-0.8cm}
\includegraphics[width=0.5\textwidth]{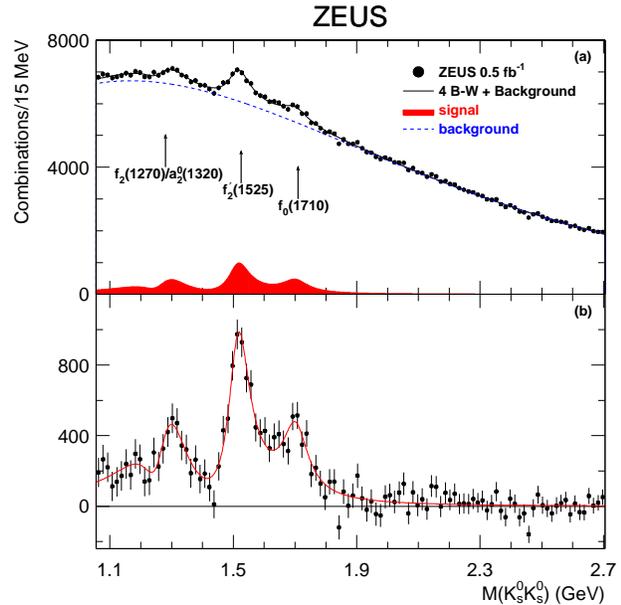}
 \vspace{-0.8cm}
\caption{The measured \ksks invariant-mass spectrum (a) and
background-subtracted \ksks invariant-mass spectrum (b) with fits}
 \label{kk}

 \vspace{-1.0cm}
 \end{center}
  \end{wrapfigure}

 The $f_{2}(1270)$, $a_{2}^{0}(1320)$
and \ffb states have $J^P = 2^+$ spin. The intensity of the
modulus-squared of the sum of these three amplitudes with ratios of
$5:-3:2$ as expected for production via an electromagnetic
process~\cite{Althoff,lipkin} and the incoherent addition of \ffc
can be derived from the SU(3) symmetry argument. Very competitive
measurements on peak position and width for \ffb and \ffc are done
with interference fit and the overall fit describes the data very
well as shown in Figure \ref{kk}. The values with statistical and
systematical uncertainties were compared well with the PDG
values~\cite{PDG}.

\section{Acknowledgments}
 I would like to thank the H1 and ZEUS collaborators for their
 efforts to produce the wonderful physics results that I presented
 on the conference. I also want to thank the conference organizers.

\end{document}